\pdfoutput=1

\documentclass[a4paper,11pt]{article}
\usepackage{jcappub} 
\usepackage{color}
\usepackage{xcolor}
\usepackage{amsmath}
\usepackage{amsfonts}
\usepackage{graphicx} 
\usepackage{amssymb}
\usepackage{epstopdf}
\usepackage{url}
\usepackage{bm}
\usepackage{hhline}
\usepackage{multirow}
\usepackage{epsfig}
\usepackage{varioref}
\usepackage{amssymb}
\usepackage{comment} 
\usepackage{cases}
\usepackage{multirow}
\usepackage{mathrsfs}
\usepackage{calrsfs}
\usepackage{mathtools}
\usepackage{amsthm}
\usepackage{hyperref}
\usepackage{subcaption}
\include{configs} 

\title{The effective two-dimensional phase space of cosmological scalar fields}
\author{David C. Edwards$^1$}
\affiliation{$^1$Institute for Astronomy, University of Edinburgh, Royal Observatory, Edinburgh EH9~3HJ, United Kingdom}

\emailAdd{dce@roe.ac.uk}
\date{\today}

\abstract{It has been shown by Remmen and Carroll \cite{remmen13} that, for a model universe which contains only a kinetically canonical scalar field minimally coupled to gravity it is possible to choose `special coordinates' to describe a two-dimensional effective phase space. The special, non-canonical, coordinates are $\phi$,$\dot{\phi}$ and the ability to describe an effective phase space with these coordinates empowers the common usage of $\phi-\dot{\phi}$ as the space to define inflationary initial conditions. This paper extends the result to the full Horndeski action. The existence of a two-dimensional effective phase space is shown for the general case. Subsets of the Horndeski action, relevant to cosmology are considered as particular examples to highlight important aspects of the procedure.}

\begin{document}
\maketitle

\section{Introduction}

Recently, Remmen and Carroll \cite{remmen13} explored the idea of considering $\phi-\dot{\phi}$ space as an effective phase space in flat Friedmann-Robertson-Walker universes. This is achieved via a many-to-one map that is \emph{vector field invariant}, that is a map which unambiguously maps a vector field on one manifold to another (lower dimensional) manifold. A rigorous definition of such a map is laid out in section 2. The importance of this map is that it preserves an important facet of the full phase space: the curves of motion do not cross. In principle, this allows for the definition of a Liouville measure on the effective phase space and so an ability to answer questions similar to that of ``How many e-folds should we expect from inflation?" \cite{remmen14}.

There are further interesting areas which depend, somewhat implicitly, on the notion of $\phi-\dot{\phi}$ being an effective phase space. In slow-roll inflation the attractor behaviour \cite{liddle94} is a ubiquitous feature which greatly simplifies the calculations. Trajectories converge towards each other in $\phi-\dot{\phi}$ in such a way that the initial value of $\dot{\phi}$ is unimportant. If this were happening in a non-phase space then there may be concerns that the convergence of trajectories did not imply a single behaviour. Furthermore, numerical investigations into inflationary behaviour, for example \cite{tegmark05} in the single-field case and \cite{frazer11,frazer12,dias12} in the multi-field case, take $\phi-\dot{\phi}$ as an appropriate space on which to define initial conditions. Again, this is only valid if it is an effective phase space.

The purpose of this paper is to extend the range of theories that the results of Remmen and Carroll apply to by showing that the full Horndeski action admits an effective phase-space description. Section 2 will deal with preliminary ideas, which are restated here to make the paper more self contained. In the next section a vector field invariant map is shown to exist for the Horndeski action via the construction of the Hamiltonian constraint and the manipulation of the components of the Hamiltonian flow vector. Finally, some examples of theories commonly considered in a cosmological setting are presented to highlight import aspects of the procedure.

\section{Preliminaries}

\subsection{Vector Field Invariant Maps and Effective Phase Spaces}

\label{sec:vectorFieldInv}

The idea of a vector field invariant map is conceptually simple. It is a map from one space to another (potentially of lower dimension) that maps the vector field unambiguously, so that there is a unique vector defined at each point. To state this in more formal terms some notation must be introduced. The space of all smooth valued functions on a manifold, $M$, is $\mathcal{F}(M)$. A pullback of a function, $f \in \mathcal{F}(N)$, by a mapping $\psi:M\rightarrow N$ is $\psi^*:\mathcal{F}(N)\rightarrow \mathcal{F}(M)$. Finally, a definition of vector field invariant map can be made. A map, $\psi:M\rightarrow N$, is a vector field invariant map with respect to a vector field, $X$, if, for any function, $f \in \mathcal{F}(N)$, and for all $q \in N$, $X_p(\psi^*f)=X_{p'}(\psi^*f)$ for all $p,p' \in \psi^{-1}(q)$.

With the concept of a vector field invariant map the question of whether a space constitutes an effective phase space is now answerable. A space, $N$, can be thought of as an effective phase space if it possible to construct a map, $\psi$, from some region of the full phase space, $M$, so that the Hamiltonian vector field is invariant under $\psi$. The key point is that the Hamiltonian vector field is tangent to the curves of motion and so if it is mapped unambiguously from $M$ to $N$ then it is uniquely defined at all points in $N$. This means that the curves of motion do not cross and so specifying a point in $N$ is enough to specify the dynamics of the system.

\subsection{The Horndeski Action}

\label{sec:hornAct}

The Horndeski action, equation (\ref{eq:HorndeskiAction}), is of interest as the most general action for a scalar-tensor theory in four dimensions that gives second-order field equations \cite{horndeski74,cliftonReview}. The requirement of second order field equations ensures that there are no Ostrogradsky instabilities \cite{ostro1850} in the theory. In this paper the DGSZ \cite{dgsz11} formulation of the theory is 
\begin{multline}
	\label{eq:HorndeskiAction}
	S_H = \int d^4x \sqrt{-g} \left(P(\phi,X)-G_3(\phi,X)\Box\phi+G_4(\phi,X)R+[(\Box\phi)^2-\phi^{;\alpha\beta}\phi_{;\alpha\beta}]G_{4X}(\phi,X)\right.
	\\\left.+G_5(\phi,X)G_{\alpha\beta}\phi^{;\alpha\beta}-\frac{1}{6}\left[(\Box\phi)^3-3(\Box\phi)\phi_{;\alpha\beta}\phi^{;\alpha\beta}+2\phi_{;\alpha\beta}\phi^{;\alpha\lambda}{\phi^{;\beta}}_{;\lambda}\right]G_{5X}(\phi,X)\right) \quad.
\end{multline}
The functions $P(\phi,X)$ and $G_i(\phi,X)$ are arbitrary, analytic functions of the scalar field, $\phi$, and $X=g^{\mu\nu}\partial_\mu\phi\partial_\nu\phi$. These arguments will not be written in the following to increase readability. Also, a subscript $\phi$ or $X$ denotes a partial derivate with respect to that quantity, semi-colons denote covariant derivatives, $\Box$ is the covariant D'Alembertian operator and greek indices run over all four spacetime dimensions.

For the purposes of this paper it is important that the action is well-posed \cite{dyer09}. This is achieved by the use of the appropriate boundary terms. In the case of General Relativity this is the Gibbons-Hawking-York term \cite{ghy77,york72}:
\begin{equation}
	B = \int d^3x\sqrt{h}K \quad,
\end{equation}
where, in this case, $h$ is the determinant of the induced metric on the boundary surface and $K$ is the extrinsic curvature of the boundary surface.

The boundary terms for the general Horndeski theory have been previously calculated by Padilla and Sivanesan  \cite{padilla13}, equations (\ref{eq:bound3}) to (\ref{eq:bound5}) and the new terms in these equations will be introduced as they are used in Section \ref{sec:vecFieldHornd}.

\begin{equation}
	\label{eq:bound3}
	B_3 = \int_{\partial N}F_3
\end{equation}
\begin{equation}
	\label{eq:bound4}
	B_4 = \int_{\partial N}2(G_4K-\left[\partial_YF_4\right]\bar{\Box}{\phi})
\end{equation}
\begin{multline}
	\label{eq:bound5}
	B_5 = \int_{\partial N}-\frac{1}{2}sG_5(K^2-{K^j}_i {K^i}_j)\phi_n\\-G_5K\bar{\Box}{\phi}+G_5{\phi^{;j}}_{;i}{K^i}_j+\frac{1}{2}\bar{R}F_5+\frac{1}{2}\left[\partial_YF_5\right]\left((\bar{\Box}{\phi})^2-{\phi^{;j}}_{;i}{\phi^{;i}}_{;j}\right)
\end{multline}

\section{A Vector Field Invariant Map of the Horndeski Action}

\label{sec:vecFieldHornd}

\subsection{The Hamiltonian Constraint Surface}

In this section the Hamiltonian constraint is explicitly constructed and the surface generated by it is shown to be independent of the scale factor of the model universe when considered in the special coordinate system of $\phi-\dot{\phi}$. As with the work carried out in \cite{remmen13} this paper is interested in the background dynamics of a spatially flat universe. To this end, systems with exact homogeneity and isotropy are considered, the Friedmann-Lemaitre-Robertson-Walker universes with no perturbations. This, in the language of quantum cosmology, is the minisuperspace approximation with a further restriction to only flat universes.

The Hamiltonian constraint arises from the time diffeomorphism invariance of the model. In the case of a flat FRW universe this invariance is encapsulated in the lapse function, $N_{\rm{lapse}}(t)$, which appears in the metric and, as such, the infinitesimal line element, equation (\ref{eq:FRWMetric}).

\begin{equation}
	\label{eq:FRWMetric}
	ds^2=-N^2_\text{lapse}(t)dt^2+a(t)^2\left[dx^2+dy^2+dz^2\right]
\end{equation}

To increase readability the lapse function, scale factor and Hubble constant will henceforth be written without their argument. The next step involves putting this metric into the Horndeski action, equation (\ref{eq:HorndeskiAction}), and ensuring that this action is well-posed. The general boundary terms given in Section \ref{sec:hornAct} simplify in the case of the minisuperspace approximation.

The boundary manifold of interest for the minisuperspace approximation is one whose normal vector is a unit vector in the time direction, $n^\nu=(-1/N_{\rm{lapse}}(t),0,0,0)$, and this fact combines with the high symmetry of the system to simplify the $B_\alpha$s considerably. The boundary is parameterised by the spacial coordinates of the four dimensional bulk and has an induced metric, $\mathbf{h},$\footnote{Symbols in bold should be understood to represent the full tensorial quantity.} given by:
\begin{equation}
	h_{ij} = g_{ij} \quad,
\end{equation}
where latin indices denote summing over the boundary coordinates (equivalently in this case, the spacial coordinates of the bulk).
Firstly, the normal derivative to the scalar field on the boundary, $\phi_n$ is given by:
\begin{equation}
	\phi_n = \left.n^\nu\partial_\nu\phi\right|_{\partial N} = \frac{\dot{\phi}}{N_{\rm{lapse}}} \quad.
\end{equation}
Examining the covariant derivative on the boundary (rather than the bulk), homogeneity gives $\phi_{;i} = 0$. From this it follows that second covariant derivatives of $\phi$ are also 0 and $Y=0$, where $Y$ is the boundary analogue of $X$. The Ricci curvature of the boundary, $\Bar{R}$, is 0 since only flat universes are of interest. Finally, $s$ depends on whether the boundary is spacelike or timelike. Here the boundary is spacelike, $s=-1$. The extrinsic curvature tensor and scalar \cite{carrollGRBook,kiefer2004quantum} are given by: 
\begin{equation}
	K_{ij}=\frac{1}{2N_{\text{lapse}}}\frac{\partial h_{ij}}{\partial t} \quad,
\end{equation}
\begin{equation}
	K = h^{ij}K_{ij}=\frac{3H}{2N_{\text{lapse}}} \quad.
\end{equation}
The final term to consider is the function $F_3$. This is defined in Ref.~\cite{padilla13} and in the case being dealt with here has the form:
\begin{equation}
	F_3 = \int^{\phi_n}_0 dx\;G_3\left(\phi,\frac{x^2}{2}\right) \quad .
\end{equation}
The main point of note is that given this definition is that $\frac{\partial F_3}{\partial\phi_n}=G_3$. Using all of this information gives the boundary terms as:
\begin{equation}
	B_3 = \int d^3x \left[a^3F_3\right] \quad,
\end{equation}
\begin{equation}
	B_4 = \int d^3x \left[\frac{3G_4a^3H}{N_{\rm{lapse}}}\right] \quad,
\end{equation}
\begin{equation}
	B_5 = \int d^3x \left[\frac{3G_5a^3H^2\dot{\phi}}{{N^3_{\rm{lapse}}}}\right] \quad.
\end{equation}
The full, well-posed, action is then:
\begin{multline}
	S = S_H-\sum_{m=3}^5B_m = \int dt \frac{a^3}{{N^5_{\rm{lapse}}}} \left[P{N^6_{\rm{lapse}}}-F_{3\phi}\dot{\phi}{N^5_{\rm{lapse}}}-(3H\dot{\phi}+6G_{4\phi}\dot{\phi}){N^4_{\rm{lapse}}}\right. \\ \left.-3H^2\left(2G_4{N^2_{\rm{lapse}}}+G_{5\phi}\dot{\phi}^2-2G_{4X}\dot{\phi}^2\right){N^2_{\rm{lapse}}}+3H^3G_{5X}\dot{\phi}^3\right] \quad.
\end{multline}

Now, the conjugate momenta can be calculated in the standard way from the Lagrangian as:
\begin{eqnarray}
	\label{eq:aMomentum}
	p_{a} & = & \frac{\partial L}{\partial\dot{\phi}}=3a^2\left[\frac{-2G_{4\phi}\dot{\phi}+1}{{N_{\rm{lapse}}}}-\frac{2H^2\left(2G_4{N^2_{\rm{lapse}}}+G_{5\phi}\dot{\phi}^2-2G_{4X}\dot{\phi}^2\right)}{{N^3_{\rm{lapse}}}}\right. \nonumber \\ & & \qquad\qquad\qquad \left.+\frac{H^2G_{5X}\dot{\phi}^3}{{N^5_{\rm{lapse}}}}\right] \\
	\label{eq:phiMomentum}
	p_{\phi} & = & \frac{\partial L}{\partial\dot{a}}=a^3\left[\frac{P_X\dot{\phi}-6HG_{4\phi}}{{N_{\rm{lapse}}}}+\frac{\dot{\phi}^2F_{3\phi X}}{N^2_{\rm{lapse}}}+\frac{6H\dot{\phi}}{{N^3_{\rm{lapse}}}}\left(HG_{4X}-HG_{5\phi}-G_{4\phi X}\dot{\phi}\right) -\frac{F_{3\phi}}{N^5_{\rm{lapse}}}\right. \nonumber \\ & & \qquad\qquad\qquad \left.+\frac{3H^2\dot{\phi}^2}{{N^5_{\rm{lapse}}}}\left(HG_{5X}+2G_{4XX}\dot{\phi}-G_{5\phi X}\dot{\phi}\right)+\frac{H^3G_{5XX}\dot{\phi}^4}{{N^7_{\rm{lapse}}}}\right] \quad.
\end{eqnarray}
Note that the momentum associated with $N_{\rm{lapse}}(t)$ is zero. This means that it is not a dynamical and can be set to an arbitrary constant, here $N_{\rm{lapse}}(t)=1$ is chosen. Since the key feature in what follows is the $a$ dependency of flow vector components it is useful to write the momenta as:
\begin{eqnarray}
	\label{eq:aMomentumFac}
	p_{a} & = & a^2\pi_a \\
	\label{eq:phiMomentumFac}
	p_{\phi} & = & a^3\pi_\phi \quad,
\end{eqnarray}
where the $\pi_i$s contain all the functional dependency of the momenta on $\phi$, $\dot{\phi}$ and $H$.
The Hamiltonian is then constructed in the usual way, equation \ref{eq:basicHamiltonian}. Since the expressions for the momenta are kept general there is no way to invert the expressions to obtain the velocities. Therefore, the Hamiltonian is written as a function of $\phi$, $\dot{\phi}$, $a$ and $H$ instead of as function of $\phi$, $p_{\phi}$, $a$ and $p_{a}$ and this also suits the purpose of this paper.
\begin{equation}
	\label{eq:basicHamiltonian}
	\mathcal{H}=\sum \dot{q_i}p_i -L
\end{equation}

The Hamiltonian constraint comes from varying the action with respect to the lapse function (before it is set to be constant), see e.g. \cite{kiefer2004quantum}, and gives the the Friedmann equation:
\begin{multline}
	H^3\dot{\phi}^3\left(G_{5XX}\dot{\phi}^2+5G_{5X}\right)-3H^2\left[\dot{\phi}^2\left(\dot{\phi}^2(G_{5\phi X}-2G_{4XX})-4G_{4X}+3G_{5\phi}\right)+2G_4\right] \\ -6H\dot{\phi}\left(G_{4\phi X}\dot{\phi}^2+G_{4\phi}\right)+\dot{\phi}^2P_X-P =  0 \quad.
\end{multline}

The two-dimensional Hamiltonian constraint surface in the space of $(\phi,\dot{\phi},H)$, $C_{a_\star}$, is the same for \emph{all} possible values of $a_\star$ so that, as in the case of Remmen and Carroll, it is found that the full constraint surface, $C$, factorises as a product: $C=C_{a_\star}\times\mathbb{R}_+$.

\subsection{The Hamiltonian Vector Field Components}

\label{sec:vectorFieldComps}

The Hamiltonian vector field is defined by:
\begin{equation}
	\label{eq:basicHamFlow}
	X_{\mathcal{H}}=\frac{\partial\mathcal{H}}{\partial p_i}\frac{\partial}{\partial q^i}-\frac{\partial\mathcal{H}}{\partial q^i}\frac{\partial}{\partial p_i} \quad.
\end{equation}
This definition lends itself to consider components in the $\phi,p_{\phi},a,p_a$ directions but the space can be described in any choice of components such as $(\phi,\dot{\phi},a,H)$. In this second coordinate system the $\dot{\phi}$ and $H$ components of the flow vector are given by:

 \begin{eqnarray}
	X_{\mathcal{H}}^{(\dot{\phi})}&=&X_{\mathcal{H}}^{(p_\phi)}\frac{\partial\dot{\phi}}{\partial p_{\phi}}+X_{\mathcal{H}}^{(p_a)}\frac{\partial\dot{\phi}}{\partial p_{a}} \quad,\label{eq:flowVectorSpecialCoordsPhiDot}\\
	X_{\mathcal{H}}^{(H)}&=&X_{\mathcal{H}}^{(p_\phi)}\frac{\partial H}{\partial p_{\phi}}+X_{\mathcal{H}}^{(p_a)}\frac{\partial H}{\partial p_{a}}\label{eq:flowVectorSpecialCoordsH} \quad.
\end{eqnarray}

In the case of the Horndeski action, the momenta are given in equations (\ref{eq:aMomentum}) and (\ref{eq:phiMomentum}) and contain the unknown functions $G_i(\phi,X)$. It is impossible to invert these expressions without specifying $G_i(\phi,X)$ and so, to keep the approach in this paper as general as possible, an alternative route must be found. Directly from Hamilton's equations the components $X_{\mathcal{H}}^{(p_\phi)}$ and $X_{\mathcal{H}}^{(p_a)}$ are given as:

\begin{eqnarray}
	X_{\mathcal{H}}^{(p_{\phi})}&= -\frac{\partial\mathcal{H}}{\partial\phi} &= \dot{p}_{\phi} = a^3\left(3H\pi_\phi+\dot{\pi}_\phi\right) \quad, \\
	X_{\mathcal{H}}^{(p_{a})}&= -\frac{\partial\mathcal{H}}{\partial a} &= \dot{p}_{a} = a^2\left(2H\pi_a+\dot{\pi}_a\right)\quad.
\end{eqnarray}
Note that the terms $\dot{\pi}_i$ will, in general, contain second derivatives of the generalised coordinates. These can be eliminated by the equations of motion which can be shown to be independent of $a$. The easiest way to see this is by writing the Lagrangian as $\mathcal{L} = a^3\lambda$. Then the equation of motion for $a$, which gives $\dot{H}$, is given by the Euler-Lagrange equations and taking derivatives with respect to $H$ instead of $\dot{a}$ gives:
\begin{equation}
	a^2\left(3\lambda -\left[2H\frac{\partial\lambda}{\partial H}+\frac{d}{dt}\frac{\partial\lambda}{\partial H}\right]\right) = 0 \quad .
\end{equation}
That is, the $a$ dependency has factored out of the equation of motion. A similar approach for the $\phi$ equation of motion yields:
\begin{equation}
	a^3\left(3\frac{\partial\lambda}{\partial\phi} -\left[3\frac{\partial\lambda}{\partial\dot{\phi}}+\frac{d}{dt}\frac{\partial\lambda}{\partial\dot{\phi}}\right]\right) = 0 \quad .
\end{equation}
Now, all that remains is to manipulate the partial derivatives terms in equations (\ref{eq:flowVectorSpecialCoordsPhiDot}) and (\ref{eq:flowVectorSpecialCoordsH}). This is a simple inversion of the expressions:
\begin{equation}
	\frac{\partial{p}_{i}}{\partial{p}_{j}}=\frac{\partial\dot{\phi}}{\partial{p}_{j}}\frac{\partial{p}_{i}}{\partial\dot{\phi}}+\frac{\partial H}{\partial{p}_{j}}\frac{\partial{p}_{i}}{\partial H} \quad.
\end{equation}
Re-writing this as a matrix equation, 
\begin{equation}
	\begin{pmatrix}\frac{\partial{p}_{\phi}}{\partial{p}_{\phi}}\\ \frac{\partial{p}_{\phi}}{\partial{p}_{a}}\\ \frac{\partial{p}_{a}}{\partial{p}_{\phi}}\\ \frac{\partial{p}_{a}}{\partial{p}_{a}}\end{pmatrix} = \begin{pmatrix}1 \\ 0 \\ 0 \\ 1 \end{pmatrix} = \bm{D}\begin{pmatrix}\frac{\partial\dot{\phi}}{\partial{p}_{\phi}}\\ \frac{\partial H}{\partial{p}_{\phi}}\\ \frac{\partial\dot{\phi}}{\partial{p}_{a}}\\ \frac{\partial H}{\partial{p}_{a}}\end{pmatrix} \quad,
\end{equation}
where $\bm{D}$ is a $4\times4$ matrix of partial derivatives, allows the flow vector components to be written in a compact form. The matrix $\bm{D}$ takes the explicit form:
\begin{equation}
	\bm{D} = \begin{pmatrix}
		\frac{\partial p_{\phi}}{\partial\dot{\phi}} & \frac{\partial p_{\phi}}{\partial H} & 0 & 0 \\
		\frac{\partial p_{a}}{\partial\dot{\phi}} & \frac{\partial p_{a}}{\partial H} & 0 & 0 \\
		0 & 0 & \frac{\partial p_{\phi}}{\partial\dot{\phi}} & \frac{\partial p_{\phi}}{\partial H} \\
		0 & 0 & \frac{\partial p_{a}}{\partial\dot{\phi}} & \frac{\partial p_{a}}{\partial H} \\
	\end{pmatrix} = \begin{pmatrix}\bm{A} & \bm{0} \\ \bm{0} & \bm{A} \end{pmatrix} \quad ,
\end{equation}
where $\bm{A}$ is now a $2\times2$ matrix. Since $\bm{D}$ takes a block form the system of equations can be written as:
\begin{eqnarray}
	\begin{pmatrix}1\\0\end{pmatrix} & = & \bm{A}\begin{pmatrix}\frac{\partial\dot{\phi}}{\partial p_{\phi}} \\ \frac{\partial\dot{H}}{\partial p_{\phi}} \end{pmatrix} \quad , \\
	\begin{pmatrix}0\\1\end{pmatrix} & = & \bm{A}\begin{pmatrix}\frac{\partial\dot{\phi}}{\partial p_{a}} \\ \frac{\partial\dot{H}}{\partial p_{a}} \end{pmatrix} \quad .
\end{eqnarray}
Given this decomposition of the equations the flow vector components can be written as:
\begin{eqnarray}
	X_{\mathcal{H}}^{(\dot{\phi})} & = & \dot{p}_{\phi}{\bm{A}^{-1}}_{1,1}+\dot{p}_{a}{\bm{A}^{-1}}_{2,1} \\
	X_{\mathcal{H}}^{(H)} & = & \dot{p}_{\phi}{\bm{A}^{-1}}_{1,2}+\dot{p}_{a}{\bm{A}^{-1}}_{2,2} \quad.
\end{eqnarray}
The matrix $\bm{A}^{-1}$ is given by:
\begin{equation}
\bm{A}^{-1} = \frac{1}{a^3\chi}\begin{pmatrix}
		\frac{\partial \pi_{a}}{\partial H} & -a\frac{\partial \pi_{\phi}}{\partial H} \\
		-\frac{\partial \pi_{a}}{\partial\dot{\phi}} & a\frac{\partial \pi_{\phi}}{\partial\dot{\phi}} 
		\end{pmatrix} \quad ,
\end{equation}
where
\begin{equation}
	\chi = \frac{\partial \pi_{\phi}}{\partial\dot{\phi}}\frac{\partial \pi_{a}}{\partial H}-\frac{\partial \pi_{\phi}}{\partial H}\frac{\partial \pi_{a}}{\partial\dot{\phi}} \quad .
\end{equation}
Despite the generality of the Horndeski action, this notation allows a clear form for the expressions of the flow vector to be written down. These are:
\begin{eqnarray}
	X_{\mathcal{H}}^{(\dot{\phi})} & = & \frac{1}{\chi} \left[H\left(3\pi_\phi\frac{\partial\pi_a}{\partial H}-2\pi_a\frac{\partial\pi_\phi}{\partial H}\right)+\dot{\pi}_\phi\frac{\partial\pi_a}{\partial H}-\dot{\pi}_a\frac{\partial\pi_\phi}{\partial H}\right] \quad,\\
	X_{\mathcal{H}}^{(H)} & = & \frac{1}{\chi} \left[H\left(2\pi_a\frac{\partial\pi_\phi}{\partial \dot{\phi}}-3\pi_\phi\frac{\partial\pi_a}{\partial \dot{\phi}}\right)+\dot{\pi}_a\frac{\partial\pi_\phi}{\partial \dot{\phi}}-\dot{\pi}_\phi\frac{\partial\pi_a}{\partial \dot{\phi}}\right] \quad.
\end{eqnarray}
That is, there is no dependence on the scale factor in the $\dot{\phi}$ and $H$ components the flow vector. This, along with the ability to write the constraint surface in terms of only $\phi$ and $\dot{\phi}$ as shown in the previous section are exactly the criteria laid out in section \ref{sec:vectorFieldInv} to consider the $\phi-\dot{\phi}$ space an \emph{effective phase space} as meant by Remmen and Carroll in Ref.~\cite{remmen13}. While the result is presented here for the Horndeski theory it follows from the argument used here that the result holds for any theory where the Friedmann equation is independent of $a$ and the momentum can be factorised as in equations (\ref{eq:aMomentumFac}) and (\ref{eq:phiMomentumFac}).

\section{Examples}

In this section specific theories that are often discussed in cosmological contexts are used to highlight interesting points that arise when considering theories more general than a minimally coupled, canonical scalar field.

\subsection{Conformally-Coupled Scalar Fields}

\label{sec:exConformal}

Theories with an action of the form:
\begin{equation}
	\label{eq:modelActionConformal}
	S=\int d^4x\sqrt{-g}\left[\frac{1}{2}\Omega(\phi)R+X+V(\phi)\right] \quad.
\end{equation}
have been considered to be viable or important in the context of inflation for several reasons. These stem from the original Starobinksy model of inflation \cite{starobinsky80} through to the more recent superconformal ideas of Kallosh and Linde \cite{kallosh13,kallosh13d} and Higgs inflation \cite{bezrukov08}. Due to this interest it makes sense to pay particular attention to important issues that arise when dealing with this set of actions in this context. These theories are a subset of the Horndeski action given in equation \ref{eq:HorndeskiAction} with $P(\phi,X)=X+V(\phi)$, $G_4(\phi,X)=\frac{1}{2}\left[1+\Omega(\phi)\right]$ and $G_3(\phi,X)=G_5(\phi,X)=0$.

Following the prescription outlined in section 3 the first key stage to arrive at is the Hamiltonian of a theory whose action is given by equation \ref{eq:modelActionConformal}. In calculating this Hamiltonian the conjugate momenta are found to be:
\begin{equation}
	\label{eq:phiMomentumConformal}
	p_{\phi}=\frac{\partial L}{\partial\dot{\phi}}=\frac{a^3\left(\dot{\phi}-3H\partial_{\phi}\Omega(\phi)\right)}{N_{\text{lapse}}} \quad,
\end{equation}
\begin{equation}
	\label{eq:aMomentumConformal}
	p_{a}=\frac{\partial L}{\partial\dot{a}}=\frac{-3a^2\left(2H\Omega(\phi)+\dot{\phi}\partial_{\phi}\Omega(\phi)\right)}{N_{\text{lapse}}} \quad.
\end{equation}
The Hamiltonian of theories of this type is then given as:
\begin{equation}
	\label{eq:HamiltonianConformal}
	\mathcal{H}=\frac{-a^3(t)\left(6\Omega(\phi)H^2+6\dot{\phi}\partial_{\phi}\Omega(\phi)H+2V(\phi){N^2_{\text{lapse}}}-\dot{\phi}^2\right)}{2N_{\text{lapse}}} \quad.
\end{equation}
The Hamiltonian constraint gives the Friedmann equation, equation (\ref{eq:FriedmannConformal}), defining a three-dimensional hypersurface in the full phase space or two-dimensional surface in the $\phi-\dot{\phi}$.
\begin{equation}
	\label{eq:FriedmannConformal}
	2V(\phi)+{\dot{\phi}}^2=6H^2\Omega(\phi)+6H\dot{\phi}\partial_{\phi}\Omega(\phi)
\end{equation}

\begin{figure}[t]
	\centering
	\includegraphics[width=0.7\textwidth,trim={10cm 0 10cm 0},clip]{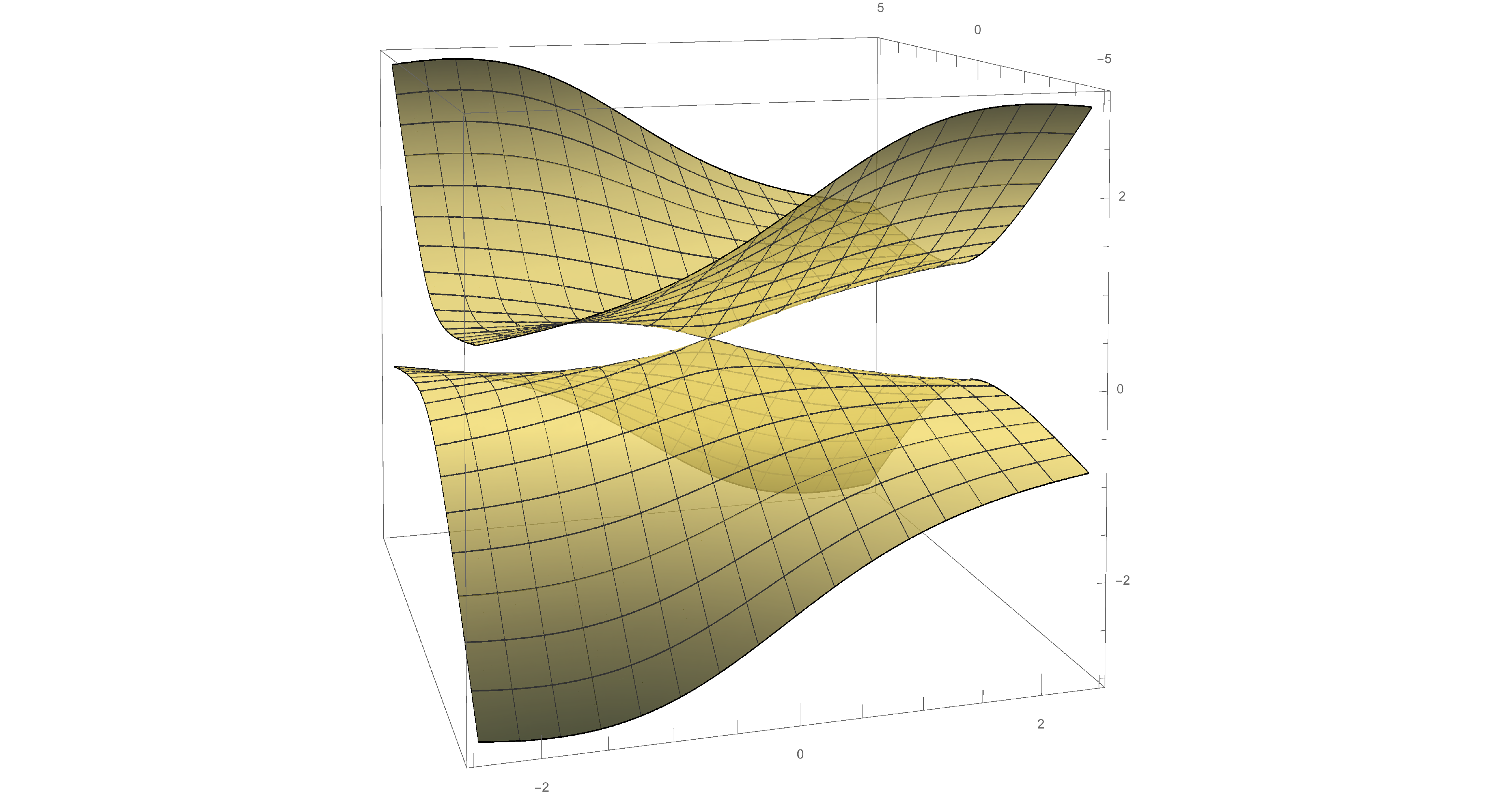}	
	\caption{A two dimensional slice, $C_{a_\star}$, of the Hamiltonian constraint surface for a model with a Lagrange density given by equation (\ref{eq:modelActionConformal}) with $\Omega(\phi)=\frac{1}{6}\phi^2$ and $V(\phi)=\frac{1}{2}\phi^2$.}
	\label{fig:hamiltonianSurface}
\end{figure}

The shape of a slice of the constraint surface for a particular coupling is shown in Figure~\ref{fig:hamiltonianSurface}. It should be noted that, due to the form of equation (\ref{eq:FriedmannConformal}) there are two solutions for $H$. This was not important in the work of Remmen and Carroll as the solutions had an exact $H\rightarrow -H$ symmetry but, as can be seen in Figure~\ref{fig:hamiltonianSurface}, the symmetry does not necessarily exist in this model. Indeed, this is generally true for any model with $G_4 \ne 1/2$ which is another reason that this example is particularly instructive. This highlights the importance of the caveat introduced in section \ref{sec:vectorFieldInv}, namely that we are mapping from some region of the full phase space, in this case the $H^+$ or $H^-$ regions.

The components of the Hamiltonian flow vector are then calculated to be:

\begin{multline}
	X_{\mathcal{H}}^{(\dot{\phi})}=\frac{3\partial_{\phi}\Omega(\phi)\left[2V(\phi)-\dot{\phi}^2+4H\dot{\phi}\partial_{\phi}\Omega(\phi))\right]}{6[\partial_{\phi}\Omega(\phi)]^2+4\Omega(\phi)}\\ -\frac{2\Omega(\phi)\left[2\partial_{\phi}V(\phi)+3H^2\partial_{\phi}\Omega(\phi)+2H\dot{\phi}\partial_{\phi}^2\Omega(\phi)\right]}{6[\partial_{\phi}\Omega(\phi)]^2+4\Omega(\phi)} \quad,
\end{multline}
\begin{multline}
	X_{\mathcal{H}}^{(H)}=\frac{6H^2[\partial_{\phi}\Omega(\phi)]^2+2H^2\Omega(\phi)+2H\dot{\phi}\partial_{\phi}\Omega(\phi)\left[3\partial_{\phi}^2\Omega(\phi)+2\right]}{6[\partial_{\phi}\Omega(\phi)]^2+4\Omega(\phi)}\\+\frac{2V(\phi)\dot{\phi}^2+2\partial_{\phi}V(\phi)\partial_{\phi}\Omega(\phi)}{6[\partial_{\phi}\Omega(\phi)]^2+4\Omega(\phi)} \quad.
\end{multline}

\begin{figure}[t]
	\includegraphics[width=0.9\textwidth]{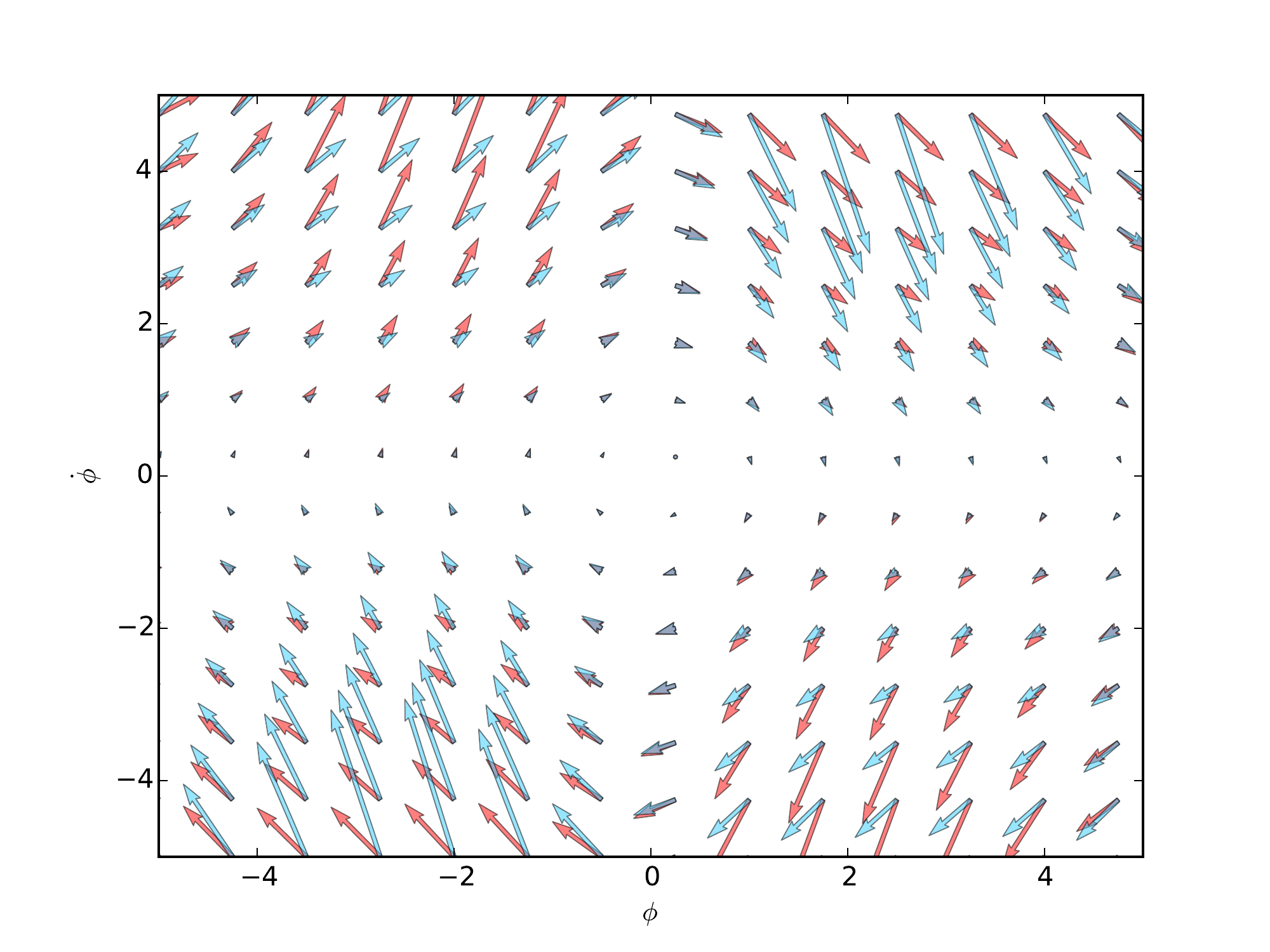}
	\caption{The two different vector fields obtained on $\phi-\dot{\phi}$ when choosing either a positive (red arrows) or negative (blue arrows) $H$ for a model with a Lagrange density given by equation (\ref{eq:modelActionConformal}) with $\Omega(\phi)=\frac{1}{6}\phi^2$ and $V(\phi)=\frac{1}{2}\phi^2$.}
	\label{fig:vectorField}
\end{figure}

The importance of the asymmetric form of the constraint surface can now be seen. Depending on whether the positive or negative solution for $H$ is chosen two different vector fields are arrived at, Figure \ref{fig:vectorField}. In many cosmological contexts, for example inflation or quintessence scenarios, this is not a problem because we choose the expanding solution. However, if the scenario of interest was a cosmology where a bounce occurs then this asymmetry in the constraint surface would suggest that the space of $\phi-\dot{\phi}$ would not constitute an effective phase space as there is not a vector field invariant map from the full region that would have to be considered. There have been explorations of the use of $\dot{\phi}-H$ as an effective phase space in simple Horndeski bounce models as it is possible to move between the positive and negative solutions \cite{Easson2011,Easson2013}.

\subsection{k-flation}

While the previous example shows one of the features that needs to be considered when dealing with more general actions it is possible to make the calculation using the exact method of Remmen and Carroll. The simplest theory that shows the need for the method presented in this paper is that of k-flation \cite{kFlation}. The action, equation (\ref{eq:modelActionKflation}), in this case contains only $P(\phi,X)$ and the usual Einstein-Hilbert term, that is $G_4(\phi,x)=\frac{1}{2}$. All the other terms from equation \ref{eq:HorndeskiAction} are zero.

\begin{equation}
	\label{eq:modelActionKflation}
	S=\int d^4x\sqrt{-g}\left[\frac{1}{2}R+P(\phi,X)\right]
\end{equation}

Once again, using the procedure of Section \ref{sec:vecFieldHornd}, the canonical momenta can be calculated and are found to be:
\begin{equation}
	\label{eq:phiMomentumKflation}
	p_{\phi}=\frac{a^3P_X\dot{\phi}}{N_{\text{lapse}}} \quad,
\end{equation}
\begin{equation}
	\label{eq:aMomentumKflation}
	p_{a}=\frac{6a^2H}{N_{\text{lapse}}} \quad.
\end{equation}
In these equations the complications that prevent the method of Remmen and Carroll being extended to more general cases than considered in Ref.~\cite{remmen13} are evident. Without knowing precisely what form $P(\phi,X)$ takes there is no way of inverting equations (\ref{eq:phiMomentumKflation}) and (\ref{eq:aMomentumKflation}) to form expressions for $\dot{\phi}$ and $H$.
\begin{equation}
	\label{eq:FriedmannKflation}
	3H^2+P=P_X\dot{\phi}^2
\end{equation}
In this case the $H\rightarrow-H$ symmetry in the constraint surface is not broken so the potential ambiguity seen in section \ref{sec:exConformal} is not present in the k-flation case. Using the method described above expressions for the flow vector components are found to be:
\begin{equation}
	\label{eq:kFlatFlowVecPhi}
	X_{\mathcal{H}}^{(\dot{\phi})}=-\frac{P_\phi}{P_X} \quad,
\end{equation}
\begin{equation}
	\label{eq:kFlatFlowVecRaw}
	X_{\mathcal{H}}^{(H)}=2H^2+\dot{H} \quad.
\end{equation}

Equation (\ref{eq:kFlatFlowVecRaw}) highlights another subtlety with this method: the appearance of second-order derivatives, in the form of $\dot{H}$, in the flow vector components. While in this case it is the second derivative of $a$ that appears, in other examples it may be $\ddot{\phi}$ that will feature. By using the equation of motion for $a$, the acceleration equation, $\dot{H}$ can be replaced. The equation of motion in this case is:
\begin{equation}
	\dot{H} = \frac{1}{2}\left(P-5H^2\right) \quad.
\end{equation}
This gives the final expression for the $H$ component of the flow vector as:
\begin{equation}
	\label{eq:kFlatFlowVecH}
	X_{\mathcal{H}}^{(H)}=\frac{1}{2}\left(P-H^2\right) \quad.
\end{equation}

Once again the flow vector components, equations (\ref{eq:kFlatFlowVecPhi}) and (\ref{eq:kFlatFlowVecH}), are independent of $a$. This, coupled with the ability to parameterise the constraint surface in terms of just $\phi$ and $\dot{\phi}$ (captured in equation \ref{eq:FriedmannKflation}) is the requirement for $\phi-\dot{\phi}$ to be considered an effective space.

\section{Conclusion}

In this paper it has been shown that the most general scalar-tensor theory with a single scalar field and second-order equations of motion, Horndeski Theory, permits a reduction to a two-dimensional effective phase space. This is an extension of a previous result of Remmen and Carroll \cite{remmen13} who showed the same for a canonical, minimally coupled scalar field. There are several features that make the general theory harder to deal with but these are elucidated through the use of specific example theories: conformally-coupled theories and k-flation.

There is still work to be done. In Ref.~\cite{remmen13} conditions are laid out for defining an effective Liouville measure on $\phi-\dot{\phi}$. Since these conditions are not proven to hold for a general potential in the simple case considered in Ref.~\cite{remmen13} no attempt has been made in this paper to show they always hold in the general case. This will need to be considered on a model by model basis. It should be noted that the measure on effective phase space is distinct from the Liouville measure on the full phase space. If it proves possible to define a Liouville measure then more sophisticated weighting could be used in numerical calculations of inflationary observables. Instead of uniform weighting in $\phi-\dot{\phi}$ uniform weighting in the Liouville measure could be used as in Ref.~\cite{remmen14}. If it can be shown to exist then this measure is the correct classical measure on effective phase space as it defined by the flow trajectories and is conserved along this flow. A flat measure on $\phi-\dot{\phi}$ is not conserved by the flow along trajectories and as such is not a valid choice.

\acknowledgments

I wish to thank Andrew Liddle for his extensive feedback and support of both the work contained in and the writing of this paper. I also wish to thank Luis Ure\~{n}a for helpful discussion relating to the slow-roll attractor behaviour of inflation. Thanks is also due to Grant Remmen for constructive feedback which greatly improved the quality of this paper. I was supported throughout this work by STFC grant number ST/K501980.

\bibliographystyle{ieeetr}
\bibliography{conformal}

\begin{thebibliography}{10}

\bibitem{remmen13}
G.~N. {Remmen} and S.~M. {Carroll}, ``{Attractor solutions in scalar-field
  cosmology},'' {\em \prd}, vol.~88, p.~083518, Oct. 2013.

\bibitem{remmen14}
G.~N. {Remmen} and S.~M. {Carroll}, ``{How many e-folds should we expect from
  high-scale inflation?},'' {\em \prd}, vol.~90, p.~063517, Sept. 2014.

\bibitem{liddle94}
A.~R. {Liddle}, P.~{Parsons}, and J.~D. {Barrow}, ``{Formalizing the slow-roll
  approximation in inflation},'' {\em \prd}, vol.~50, pp.~7222--7232, Dec.
  1994.

\bibitem{tegmark05}
M.~{Tegmark}, ``{What does inflation really predict?},'' {\em \jcap}, vol.~4,
  p.~001, Apr. 2005.

\bibitem{frazer11}
J.~{Frazer} and A.~R. {Liddle}, ``{Exploring a string-like landscape},'' {\em
  \jcap}, vol.~2, p.~026, Feb. 2011.

\bibitem{frazer12}
J.~{Frazer} and A.~R. {Liddle}, ``{Multi-field inflation with random
  potentials: field dimension, feature scale and non-Gaussianity},'' {\em
  \jcap}, vol.~2, p.~039, Feb. 2012.

\bibitem{dias12}
M.~{Dias}, J.~{Frazer}, and A.~R. {Liddle}, ``{Multifield consequences for
  D-brane inflation},'' {\em \jcap}, vol.~6, p.~020, June 2012.

\bibitem{horndeski74}
G.~W. Horndeski, ``Second-order scalar-tensor field equations in a
  four-dimensional space,'' {\em International Journal of Theoretical Physics},
  vol.~10, no.~6, pp.~363--384.

\bibitem{cliftonReview}
T.~{Clifton}, P.~G. {Ferreira}, A.~{Padilla}, and C.~{Skordis}, ``{Modified
  gravity and cosmology},'' {\em \physrep}, vol.~513, pp.~1--189, Mar. 2012.

\bibitem{ostro1850}
M.~Ostrogradski {\em Mem. Acad. St. Petersbourg}, vol.~VI, 1850.

\bibitem{dgsz11}
C.~{Deffayet}, X.~{Gao}, D.~A. {Steer}, and G.~{Zahariade}, ``{From k-essence
  to generalized Galileons},'' {\em \prd}, vol.~84, p.~064039, Sept. 2011.

\bibitem{dyer09}
E.~{Dyer} and K.~{Hinterbichler}, ``{Boundary terms, variational principles,
  and higher derivative modified gravity},'' {\em \prd}, vol.~79, p.~024028,
  Jan. 2009.

\bibitem{ghy77}
G.~W. {Gibbons} and S.~W. {Hawking}, ``{Action integrals and partition
  functions in quantum gravity},'' {\em \prd}, vol.~15, pp.~2752--2756, May
  1977.

\bibitem{york72}
J.~W. York, ``Role of conformal three-geometry in the dynamics of
  gravitation,'' {\em Phys. Rev. Lett.}, vol.~28, pp.~1082--1085, Apr 1972.

\bibitem{padilla13}
A.~{Padilla} and V.~{Sivanesan}, ``{Boundary terms and junction conditions for
  generalized scalar-tensor theories},'' {\em Journal of High Energy Physics},
  vol.~8, p.~122, Aug. 2012.

\bibitem{carrollGRBook}
S.~M. {Carroll}, {\em {Spacetime and geometry. An introduction to general
  relativity}}.
\newblock 2004.

\bibitem{kiefer2004quantum}
C.~Kiefer, {\em Quantum Gravity}.
\newblock International series of monographs on physics, Clarendon Press, 2004.

\bibitem{starobinsky80}
A.~A. {Starobinsky}, ``{A new type of isotropic cosmological models without
  singularity},'' {\em Physics Letters B}, vol.~91, pp.~99--102, Mar. 1980.

\bibitem{kallosh13}
R.~{Kallosh}, A.~{Linde}, and D.~{Roest}, ``{Universal Attractor for Inflation
  at Strong Coupling},'' {\em Physical Review Letters}, vol.~112, p.~011303,
  Jan. 2014.

\bibitem{kallosh13d}
R.~{Kallosh} and A.~{Linde}, ``{Universality class in conformal inflation},''
  {\em \jcap}, vol.~7, p.~2, July 2013.

\bibitem{bezrukov08}
F.~{Bezrukov} and M.~{Shaposhnikov}, ``{The Standard Model Higgs boson as the
  inflaton},'' {\em Physics Letters B}, vol.~659, pp.~703--706, Jan. 2008.

\bibitem{Easson2011}
D.~A. Easson, I.~Sawicki, and A.~Vikman, ``{G-Bounce},'' {\em JCAP}, vol.~1111,
  p.~021, 2011.

\bibitem{Easson2013}
D.~A. Easson, I.~Sawicki, and A.~Vikman, ``{When Matter Matters},'' {\em JCAP},
  vol.~1307, p.~014, 2013.

\bibitem{kFlation}
C.~{Armend{\'a}riz-Pic{\'o}n}, T.~{Damour}, and V.~{Mukhanov},
  ``{k-Inflation},'' {\em Physics Letters B}, vol.~458, pp.~209--218, July
  1999.

\end{thebibliography}

\end{document}